\documentclass[apl,twocolumn,preprintnumbers,amsmath,amssymb,floatfix]{revtex4-2}
\usepackage{graphicx}
\usepackage{lmodern}
\usepackage{tabularx}
\usepackage{siunitx}
\usepackage{booktabs}
\usepackage{etoolbox}
\usepackage{epstopdf}
\usepackage{float}
\usepackage{placeins}
\usepackage{hyperref,graphicx}
\usepackage{dcolumn}
\usepackage{bm}
\usepackage{color}
\usepackage{natbib}
\usepackage{amsmath}
\usepackage{amssymb}
\usepackage{multirow} 
\usepackage{hyperref}
\usepackage{stackengine}
\hypersetup{colorlinks=true, linkcolor=blue, citecolor=red, urlcolor=magenta, pdftitle={DDPOs}, pdfauthor={Saurabh Ghosh}}
\begin{document}
\title{Design of Magnetic Polar Double-Double Perovskite Oxides through Cation Ordering} 
\author{Monirul Shaikh$^a$}
\email{msk.phe@gmail.com}
\author{Duo Wang$^b$}
\author{Saurabh Ghosh$^a$}
\email{saurabhghosh2802@gmail.com}
\affiliation{$^a$Department of Physics and Nanotechnology, SRM Institute of Science and Technology, Kattankulathur 603 203, Chennai, India}
\affiliation{$^b$Faculty of Applied Sciences, Macao Polytechnic University, Macao SAR 999078, China}
\begin{abstract}
Commencing from the centrosymmetric MnRMnSbO$_6$ compound, we explore the realm of magnetic polar double-double perovskite oxides characterized by significant ferroelectric polarization. Employing symmetry operations, first-principles methodologies, and Monte Carlo simulations, our investigation delves into the structural, magnetic, ferroelectric, and electronic attributes of the polar LaFeMnNiO$_6$ and LaTiMnNiO$_6$ compounds. The structural analysis uncovers that the paraelectric-ferroelectric phase transition is intricately linked to the Fe/Ti-displacement of square planar Fe/TiO$_4$. Notably, the magnetic LaFeMnNiO$_6$ and LaTiMnNiO$_6$ compounds demonstrate robust ferroelectric polarizations, measuring 20.0 $\mu$C/cm$^2$ and 21.8 $\mu$C/cm$^2$, respectively, accompanied by minimalist forbidden energy gaps of 1.40 eV and 1.18 eV using the GGA+U method. Furthermore, we pinpoint elevated magnetic transition temperatures for these compounds.
Additionally, our study scrutinizes the energies associated with diverse spin configurations and identifies potential minimum decomposition pathways into stable oxides. This comprehensive analysis ensures the meticulous formation of the LaFeMnNiO$_6$ and LaTiMnNiO$_6$ compounds.
\end{abstract}
\maketitle
\section{INTRODUCTION}
Double perovskite oxides (DPOs) with common chemical formulae A$_2$BB$^{\prime}$O$_6$ have intense research focus due to their implications in sciences and in modern-day technologies. They exhibit important physical properties, including ferroelectricity, ferro/ferrimagnetism, and so on \cite{tokura2014multiferroics, matsukura2015control, mundy2016atomically, fiebig2016evolution, spaldin2019advances}. In DPOs family a framework of corner-shared BO$_6$ octahedra forms a three-dimensional network and large A-site cations are accommodated at the 12-coordinated cuboctahedral cavities \cite{belik2018rise}. In addition, the A- and B-sublattices can be ordered in various ways, namely layer, columnar, and rock-salt. There are a good number of perovskite oxides that reportedly show large ferroelectric distortions with high polarization values but with negligible magnetization values \cite{PhysRevB-SG-2015, MonirulPhysRevB}.
\par
 Combining cation orderings with chemical substitution in the perovskite oxides family, is worth exploring for multiferroic properties where both ferroelectricity, and magnetism can coexist \cite{spaldin2005renaissance, eerenstein2006multiferroic, ramesh2007multiferroics}.
DPOs AA$^{\prime}$BB$^{\prime}$O$_6$ where A, and A$^{\prime}$ are alkali metals or rare-earth ions and B, and B$^{\prime}$ are transition metals are predicted and extensively studied for multiferroic properties, where both ferroelectricity and magnetism can coexist~\cite{Kanzig1957, PhysRevLett-Cochran, spldnfb, scottrev, ME0}. Due to structural and compositional flexibility in AA$^{\prime}$BB$^{\prime}$O$_6$ DPOs, both A- and B-sublattices can host a large number of atomic combinations. Recently, we realized the importance of cation ordering in both A- and B-sublattices to design polar magnetic metals/insulators in the DPOs family~\cite{CmemMat2021}. Leveraging machine learning, we have pinpointed key features that contribute to stabilizing the system, leading to A-site layered B-site rocksalt ordering~\cite{ghosh2022insights}. Additionally, our investigation has revealed that the emergence of ferroelectricity is attributed to an incommensurate cation radius mismatch between successive AO and A$^\prime$O layers (within A-site layered ordering) and structural distortion responsible for ferroelectric switching in DPOs~\cite{GPChem2024}.
\par
Double-double perovskite oxides (DDPOs) are characterized by the same formula as DPOs, i.e., AA$^{\prime}$BB$^{\prime}$O$_6$, but now are occupied by transition metals (TMs), represent a potential class of materials where functional properties including multiferroicity can be explored~\cite{aimi2014high, solana2016double, gou2017site, li2018new, ji2023cafefenbo}. 
A unique feature of these perovskite oxides is that the different atomic sites i.e., A$^{\prime}$, B and B$^{\prime}$ can accommodate magnetic transition metals with the possibility of enhanced magnetic interactions \cite{solana2016double, ji2023cafefenbo}. 
However, the first reported DDPO is CaFeTi$_2$O$_6$ with only one magnetic (Fe) site~\cite{leinenweber1995high}. 
It crystallizes in a tetragonal centrosymmetric space group \textit{P4$_2$/nmc} with a 10-fold Ca-coordination, a tetrahedral Fe-coordination, and another square-planer Fe-coordination. 
In recent years, A-site ordered CaMnTi$_2$O$_6$ DDPOs with similar coordination of CaFeTi$_2$O$_6$ at the A-sublattices, attracted attention owing to its ferroelectric distortion with large bandgap $\sim$ 3.0 eV ~\cite{aimi2014high, gou2017site, li2018new}. 
A set of similar DDPOs, MnRMnSbO$_6$ (R = La, Pr, Nd, Sm) with both A-site and B-site cation ordered phases are synthesized \cite{solana2016double}. 
These DDPOs show large magnetization but they crystallize in tetragonal centrosymmetric space group $P4_2/n$ with no ferroelectric polarization \cite{solana2016double, ji2023cafefenbo}. 
\par
In addition to the cation orderings at the A- and B-sublattices, 15 different tilting patterns lower the symmetry from cubic \textit{Pm$\bar{3}$m} and give rise to various fascinating physical properties  ~\cite{reedijk2013comprehensive}. Out of which a few Glazer patterns such as $a^+a^+a^+$, $a^0b^+b^+$, $a^0b^+b^-$, $a^-a^-c^+$ have been explored for A-site cation ordered DPOs with in-phase (+), out-of-phase (-), and no (0) rotations of the BO$_6$ octahedra ~\cite{howard1998group, li2018spin}. 
The $a^+a^+c^-$ rotational pattern \cite{gou2017site, Glazera09401} with A/A$^{\prime}$ cation ordered DDPOs can lead to ferroelectric distortion into the structure.
Thus, the B-site rocksalt A-site columnar double-double perovskite oxides with appropriate rotational patterns hold promise for fabricating magnetic materials with substantial ferroelectric polarization. 
\par
In this work, our focus lies on the deliberate design of magnetic ferroelectrics. We aim to achieve this by introducing $a^+a^+c^-$ octahedra rotations within MnRMnSbO$_6$ double-double perovskite oxides (DDPOs). This design strategy involves cation ordering at both A- and B-sublattices, coupled with chemical substitution. Within first principles framework, we discuss the origin of ferroelectricty, band gaps opening, and high magnetic transition temperatures in polar LaFeMnNiO$_6$ (LFMNO) and LaTiMnNiO$_6$ (LTMNO) compounds. We study the stability of these compounds for different spin configurations, and possible minimum decomposition pathways into the stable oxides. 
\section{COMPUTATIONAL METHODS}
Density functional theory (DFT) ~\cite{DFT} calculations are performed for optimization of geometry, total energy, and polarization using the Berry phase method ~\cite{Berry} as implemented in the Vienna $ab-initio$ simulation package (VASP) ~\cite{vasp}.
The k-integration in the Brillouin zone is incorporated using $\Gamma$-centered 4 × 4 × 4 points for geometry optimization and 8 × 8 × 8 points for self-consistent calculations using 520.0 eV energy cut-off. We considered the generalized gradient approximation (GGA) augmented by the Hubbard-$U$ corrections (GGA+$U$) ~\cite{LSDAU} to describe the exchange-correlation effect. To consider $d-d$ Coulomb interactions, we employ $U_E$ ~\cite{Dudarev1} (= $U-J_H$ where $J_H$ is Hund’s exchange parameter) parameters of 0.0 eV for Ti-$d$ ~\cite{gou2017site}, 4.0 eV for both Mn-$d$, and Fe-$d$, and 6.0 eV for Ni-$d$ ~\cite{CmemMat2021} electrons. The exchange-correlation part is estimated by Perdew-Burke-Ernzerhof revision for solids (PBEsol) functional ~\cite{pbesol}. The total energy and Hellman-Feynman force are carefully converged for individual atoms down to 1 $\mu eV$ and 1 meV/$\AA$, respectively. 
To draw, and analyze the geometry of our three-dimensional systems, we implement the Visualization for Electronic and Structural Analysis (VESTA) software \cite{vesta1}. We perform phonon calculations on the fully relaxed $a^+a^+c^-$ rotated LaTMMnNiO$_6$ DDPOs to find polar structural distortion using the finite difference method as implemented in VASP \cite{finite1}. The symmetry operations are performed with the help of the ISODISTORT tool ~\cite{isodisplace, campbell2006isodisplace1}.
\par
Next, we utilize the optimized structure as input for the calculations of interatomic exchange parameters by means of the magnetic force theorem (MFT) ~\cite{10.1016/0304-8853(87)90721-9} using full-potential linear muffin-tin orbital (FP-LMTO) in RSPT code ~\cite{wills2010full}. At last, an effective spin Hamiltonian is constructed, and phase transition temperatures are obtained by conducting classical Monte Carlo simulations, as implemented in the UppASD package ~\cite{eriksson2017atomistic}. Exchange parameters are calculated using the full-potential linear muffin-tin orbitals code RSPt ~\cite{wills2010full}, and these parameters are then used as input for classical Monte Carlo simulations, as implemented in the UppASD package ~\cite{eriksson2017atomistic}. Supercells consisting of approximately 60,000 magnetic atoms are adopted as the structural model. An annealing process is simulated by performing calculations that start at high temperatures and are gradually decreased to 0 K. To ensure that the magnetic properties obtained at each temperature are in their equilibrium state, an initial simulation of 50,000 steps is performed.
\par
We calculate the formation energies for each DDPOs proposed in this work by considering the decomposition reaction of the DDPOs via the most probable reaction pathways. Many possible reaction sequences can produce these DDPOs, and each of them could lead us to different formation energies, which could give erroneous results. Hence, to find the correct formation energy, we find out the minimum energy reaction sequence to produce these DDPOs.  
\section{RESULTS AND DISCUSSIONS}
\begin{figure}
\centering
\includegraphics[width=\linewidth]{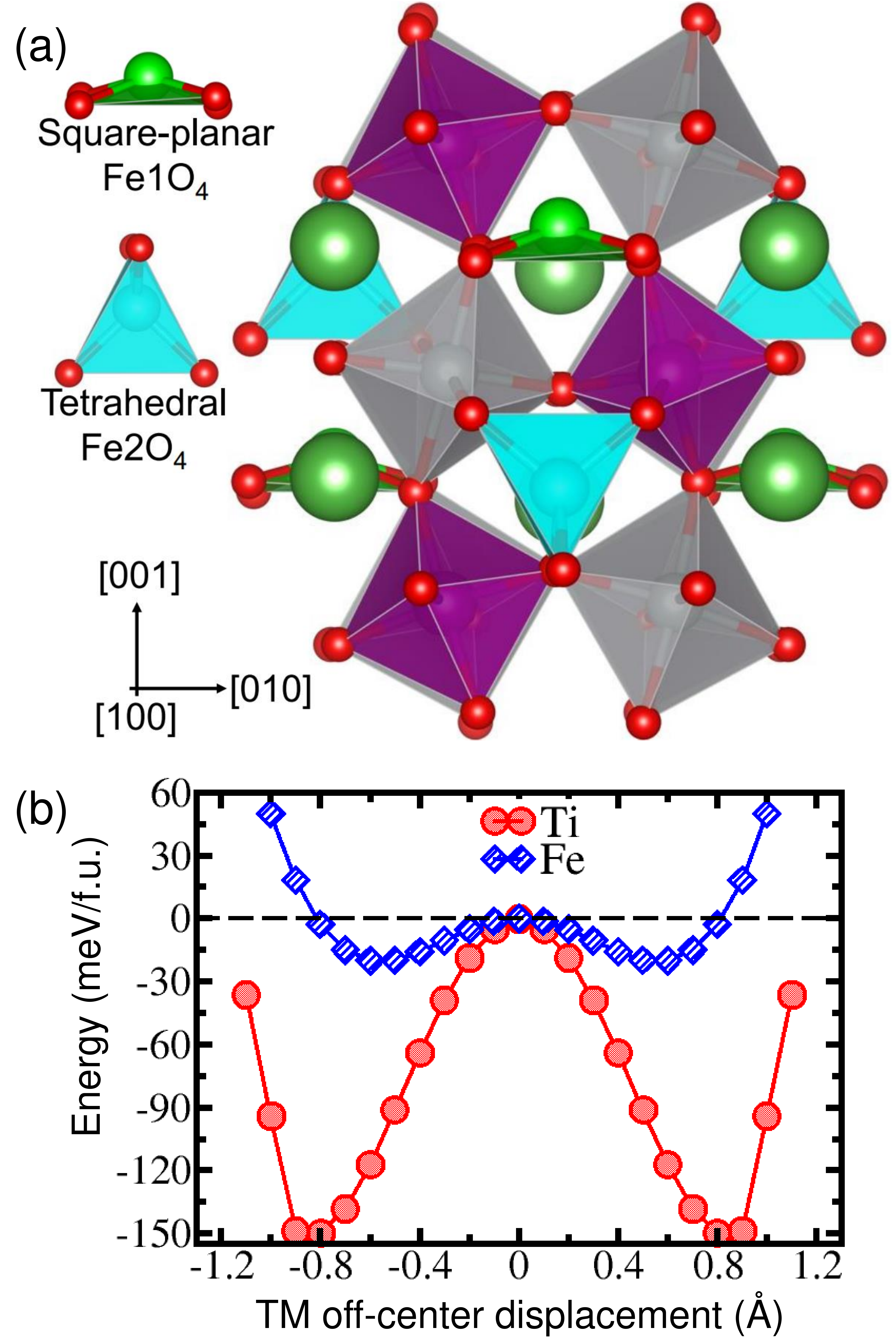}\vspace{-0pt}
\caption {(Color online) (a) Crystal structure of ferroelectric LaFeMnNiO$_6$ DDPO. B-site NiO$_6$ and MnO$_6$ octahedra are arranged in a rock-salt type order whereas A-site La and Fe are ordered in columns parallel to the crystallographic $c$-direction. Further, Fe A-sites govern Fe1O$_4$ and Fe2O$_4$ with square-planar and tetrahedral environments respectively. The La-, Mn-, Ni-, and O-atoms are described by green, magenta, silver, and red balls, respectively. (b) The double-well potential of LFMNO(blue rhombus) and LTMNO(red circles) as a function of Fe/Ti off-center displacements from the square planar environment as is shown in Figure \ref{Structures} (a).} 
\label{Structures}
\end{figure}
\begin{table}[b]
\caption{\label{tab:2} Intriguing physical properties of LFMNO and LTMNO DDPOs.}
\begin{ruledtabular}
\begin{tabular}{cccccc}
 Systems & Space  & Band gap& Polarization & Magnetization & T$_C$\\
         & group & E$_g$ (eV)&$\mu C$/cm$^2$ & $\mu_B$/f.u. & (K)  \\
\hline
LFMNO & $P4_2$ & 1.40 &20.0 & 5.0 & 225\\
LTMNO & $P4_2$ & 1.18 &21.8 & 0.0 & 48 \\
\end{tabular}
\end{ruledtabular}
\end{table}
\begin{table}[b]
\caption{\label{tab:3} Crystal structure information of LaTMMnNiO$_6$ with TM = Fe and Ti.}
\begin{ruledtabular}
\begin{tabular}{ccccc}
 Systems &    & Lattice parameters ($\AA$) &   & Cell \\
         &  $a$  & $b$ & $c$ & volume ($\AA^3$) \\
\hline
LFMNO & 7.55 & 7.55 & 7.74 & 441.20 \\
LTMNO & 7.65 & 7.65 & 7.76 & 454.44 \\
\end{tabular}
\end{ruledtabular}
\end{table}
\begin{figure*}
\centering
\includegraphics[height=4.5 cm]{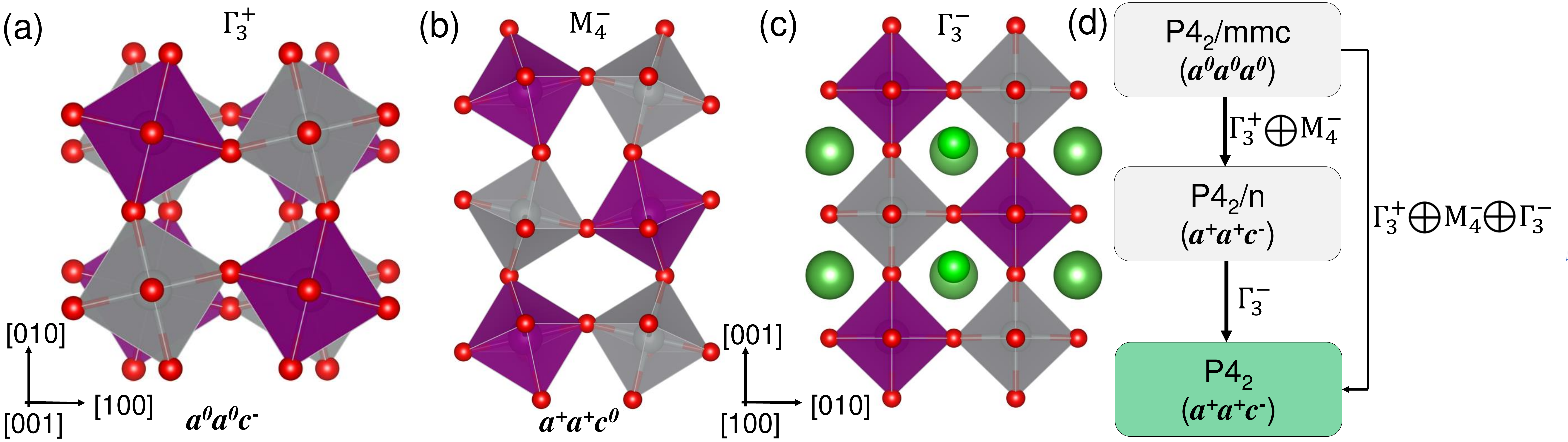}\vspace{-0pt}
\caption {(Color online) The low symmetry ferroelectric $P4_2$ phase of LaTMMnNiO$_6$ compounds for $a^+a^+c^-$ rotation is related to the centrosymmetric $P4_2/mmc$ reference structure through three major structural distortions; (a) out-of-phase ($a^0a^0c^-$) rotation of Fe/TiO$_6$ octahedra along $c$-axis denoted by $\Gamma_3^+$ irreducible representation (irrep.), (b) in-phase ($a^+a^+c^0$) rotation of Fe/TiO$_6$ octahedra along two crystallographic $a$- and $b$-axes represented by M$_4^-$ irrep. and (c) ferroelectric displacement of the Fe/Ti-cations from the square plane of Fe/TiO$_4$ described by $\Gamma_3^-$. (d) The group-subgroup tree for our systems, the ferroelectric $P4_2$ phase, is highlighted with blue. The La-, and Fe/Ti-atoms are omitted from the first two structures for clarity.} 
\label{GroupSubgroup}
\end{figure*}
\begin{figure*}
\centering

\includegraphics[height=9.5 cm]{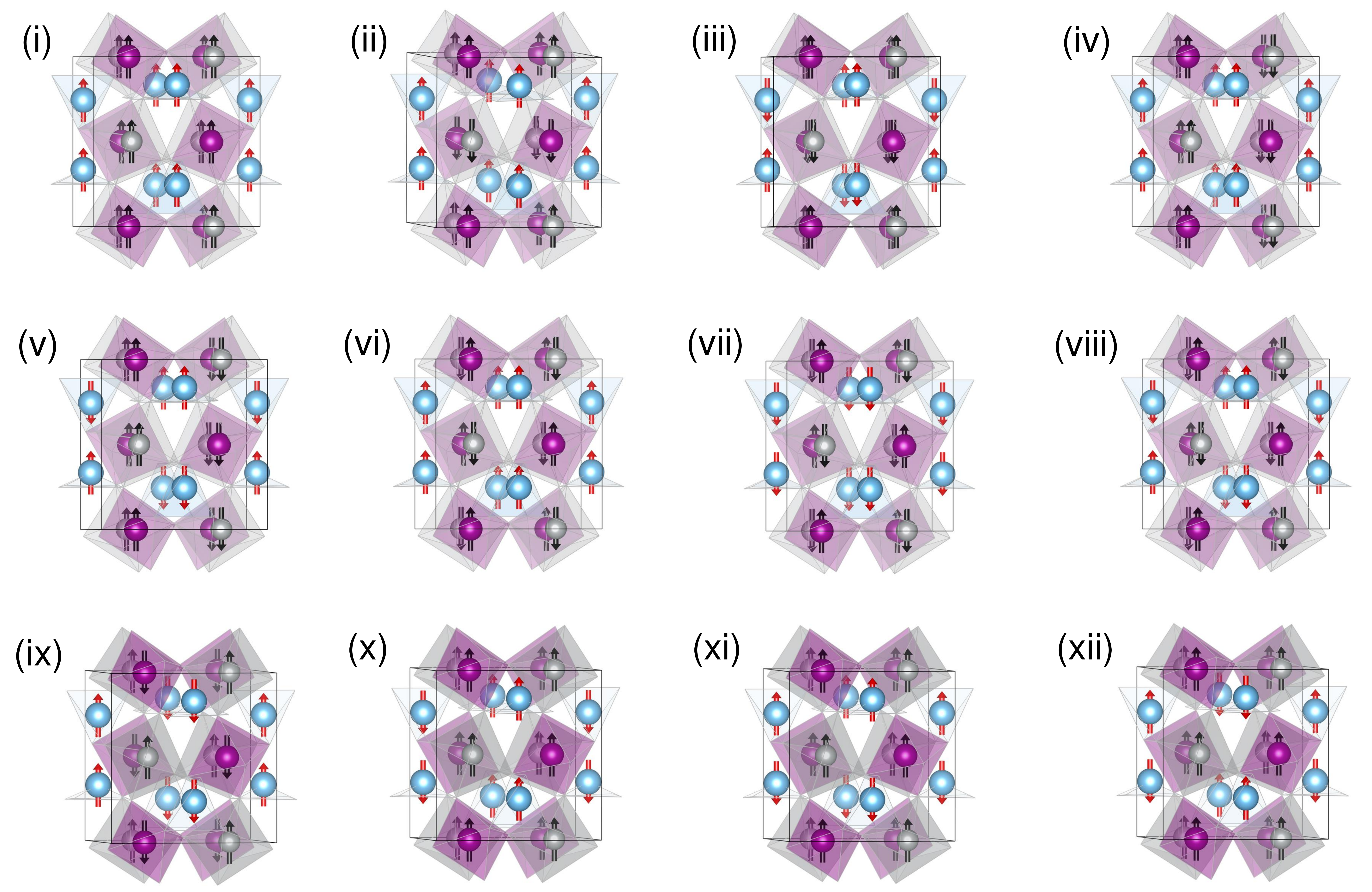}\vspace{-0pt}
\caption {(Color online) All possible spin configurations of LaFeMnNiO$_6$ within the collinear picture. The non-magnetic atoms are not shown for better clarity. The Fe-, Mn-, and Ni- atoms are denoted by blue, magenta, and silver balls respectively.} 
\label{FeMagnetism}
\end{figure*}
\begin{table*}
\caption{\label{tab:4} Magnetic moments and charge states of LaTMMnNiO$_6$ DDPOs.}
\begin{ruledtabular}
\begin{tabular}{cccccccc}
 Systems &    & Magnetic moments ($\mu_B$) of &    & Total magnetic moment &    &  Charge state of &    \\
         &  A$^{\prime}$-site  & B-site & B$^{\prime}$-site &  ($\mu_B$/f.u.)  &A$^{\prime}$-site  & B-site & B$^{\prime}$-site \\
\hline
LFMNO & 4.21 & 3.16 & 1.71  & 5.00 & 3+ & 4+ & 2+\\
LTMNO & 0.00 & 3.82 & 1.74 & 0.00 & 4+ & 3+ & 2+\\
\end{tabular}
\end{ruledtabular}
\end{table*}
\begin{figure*}
\centering

\includegraphics[height=3 cm]{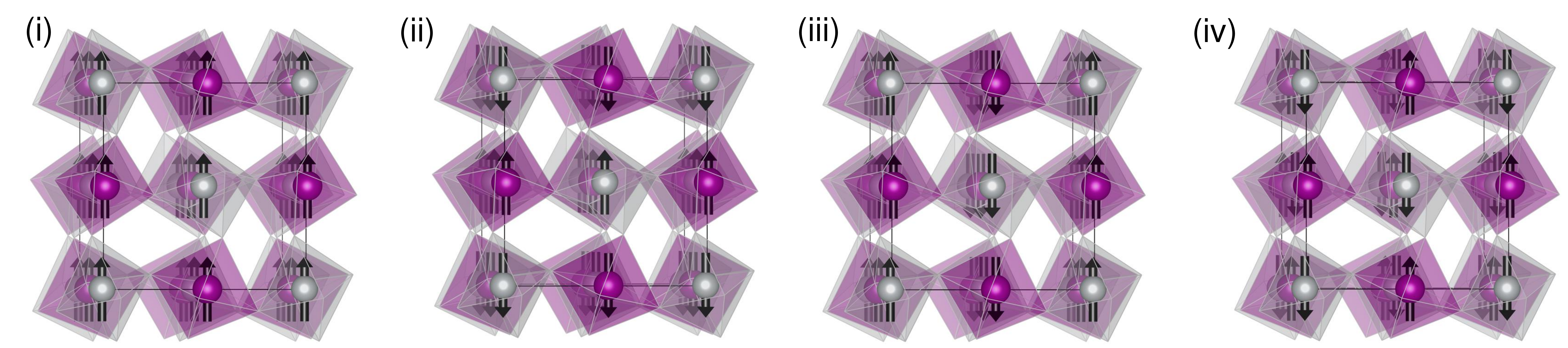}\vspace{-0pt}
\caption {(Color online) All possible spin configurations of LaTiMnNiO$_6$ within the collinear picture. The non-magnetic atoms are not shown for better clarity. The Mn-, and Ni- atoms are denoted by magenta, and silver balls respectively.} 
\label{TiMagnetism}
\end{figure*}
\begin{table*}
\begin{center}
\caption{\label{tab:5}Bond angles, and bond lengths of LaTMMnNiO$_{6}$ DDPOs.}
\begin{ruledtabular}
\begin{tabular}{ccccccc}
Systems & \multicolumn{2}{c}{Average bond angle ($^\circ$)} & \multicolumn{4}{c}{Average bond length (\r{A})} \\
        & O-TM$_{SP}$-O & Mn-O-Ni & SP TM-O  & TH TM-O & Mn-O & Ni-O\\ 
\hline
LFMNO & 86.8 &  141.7 & 2.02 & 1.94 & 1.96 & 2.03\\
LTMNO & 84.0 & 139.9 & 1.98 & 1.87 & 2.02 & 2.04\\
\end{tabular}
\end{ruledtabular}
\end{center}
\end{table*}
\subsection{Structural and Ferroelectric Properties}
In our pursuit of notable magnetic polar materials, specifically LaFeMnNiO$_6$ (LFMNO)and LaTiMnNiO$_6$ (LTMNO), we have undertaken an exploration of a series of double-double perovskite oxides (DDPOs) based on 3$d$ transition metals. These materials feature A$^{\prime}$, representing Ti, V, Cr, Mn, Fe, Co, Ni, and Cu in LaA$^{\prime}$MnNiO$_6$ type perovskites. The crystal structure of the A-site ordered DDPO is illustrated in Figure~\ref{Structures}(a), adopting a tetragonal polar (non-centrosymmetric) $P4_2$ symmetry and displaying a substantial ferroelectric polarization (see TABLE \ref{tab:2}). A detailed structural and symmetry analysis reveals that the polar P4$_2$ space group follows  
$a^+a^+c^-$ (Fe/Ti)O$_6$ octahedra rotation pattern akin to the $P4_2/nmc$ structure observed in the first-synthesized CaFeTi2O$_6$ DDPO ~\cite{leinenweber1995high}. 
Recently, Ji et al. suggested that A-site columnar and B-site rocksalt-ordered DDPOs result in a centrosymmetric $P4_2/n$ space group~\cite{ji2023cafefenbo}. However, in contrast to their findings, we present A-site columnar and B-site rocksalt-ordered DDPOs exhibiting a polar $P4_2$  structure. Notably, our design accommodates magnetic transition metals (TM) at both B- and B$^\prime$-sublattices, in addition to TM at the A$^\prime$ site, adding a layer of versatility to the system.
The La-cations acquire a 10-coordinated geometry, while both Mn- and Ni- at the B-sites form BO$_6$ octahedra almost equally tilted along the crystallographic $c$-axis. The TMs on the other hand at A$^\prime$-site show two nonequivalent geometry in columns, i.e. (Fe/Ti)O$_4$ tetrahedra (TH), and (Fe/Ti)O$_4$ square planar (SP) as are shown in cyan and green respectively in Figure~\ref{Structures} (a). Further, the space group $P4_2$ allows both Fe$^{3+}$ ($d^5$) and Ti$^{4+}$ ($d^0$) at the SP-site to move along the crystallographic $c$-direction, leading to breaking of an inversion center similar to ref. \cite{aimi2014high, gou2017site}. The absence of a center of symmetry steers ferroelectricity into the systems. Detailed information on crystal structures, magnetic moments, and charge states of these compounds are provided in TABLE \ref{tab:3}-\ref{tab:5}. Nonetheless, total energy as a function of polar distortion for Ti$^{4+}$ ($d^0$) in LaTiMnNiO$_6$ is comparable with the PbTiO$_3$ \cite{zhang2017comparative}. While Fe$^{3+}$ ($d^5$) in LaFeMnNiO$_6$ exhibit a pronounced depth of the double-well as compared to CaMnTi$_2$O$_6$ as shown in Figure~\ref{Structures} (b) \cite{gou2017site}. 
\par
We then examine the group-subgroup relation that connects between $P4_2$ 
(No. 77) phase with reference $P4_2/mmc$ (No. 131) high symmetry structure (without any distortions), and by implementing  ISODISTORT~\cite{stokes1991group}. The decomposition of the $P4_2$ phase with reference to $P4_2/mmc$ symmetry provides us with three contributing structural distortions. These are out-of-phase rotated $a^0a^0c^-$ (Fe/Ti)O$_6$ octahedra with irreducible representation (irrep.) $\Gamma_3^+$, in-phase $a^+a^+c^0$ rotation of (Fe/Ti)O$_6$ octahedra with irrep. M$_4^-$, and off-centering of (Fe/Ti)-cation of (Fe/Ti)O$_4$ SP from the center-of-symmetry with irep. $\Gamma_3^-$ as shown in Figure \ref{GroupSubgroup} (a-c). The normalized mode amplitudes 0.34 (0.36) $\AA$, 0.50 (0.53) $\AA$, and 0.16 (0.11) $\AA$ for $\Gamma_3^+$, M$_4^-$, and $\Gamma_3^-$ respectively of LTMNO (LFMNO) explain the depth of the double-wells in Figure~\ref{Structures} (b). Moreover, symmetry operation indicates a phase-transition chain described in Figure \ref{GroupSubgroup} (d). A coupling between in-phase $a^+a^+c^0$ ($\Gamma_3^+$), and out-of-phase $a^0a^0c^-$ (M$_4^-$) rotated (Fe/TiO)$_6$ octahedra reduces the symmetry to centrosymmetric $P4_2/n$ (No. 86) phase. Further reduction in symmetry occurs to polar $P4_2$ by clubbing $\Gamma_3^-$ mode into $P4_2/n$ phase.       
\subsection{Polar Magnetic Behaviors from Electronic Structure Calculations}
Next, we investigate the stability of the working compounds from various spin configurations. In the case of La$^{3+}$Fe$^{3+}$Mn$^{4+}$Ni$^{2+}$O$_6$ DDPO, the A$^\prime$-site Fe-atoms adapted within the cavity of (Mn/Ni)O$_6$ octahedra. Herein, the nearest-neighbor distances reduce significantly. Consequently, the structure exhibits a complex magnetism within the collinear spin configuration. To achieve a proper magnetic ground state in the collinear spin configuration for LFMNO system, we consider all possible spin configurations between Fe$^{3+}$, Mn$^{4+}$, and Ni$^{2+}$ and are described in Figure \ref{FeMagnetism}.  Out of which a complex ferrimagnetic configuration is found to be the magnetic ground state as shown in Figure \ref{DOS} (a). The corresponding electronic structure is shown in Figure \ref{DOS} (b). Other collinear spin configurations are found to be stable within an energy window of $\sim$ 1 eV. A set of similar complex ferrimagnetic compounds were synthesized previously but all of them were found to be centrosymmetric $P4_2/n$ space group \cite{mcnally2017complex}. All the nearest neighbor Fe spins are aligned in opposite directions, leading to a net zero moment from the Fe site in the system. Thus, Fe-Fe interactions are antiferromagnetic. 
An weak AFM Fe$^{3+}$-O-O-Fe$^{3+}$ superexchange interaction is found in Fe$^{3+}$ layers. The Fe$^{3+}$ ($d^5$) displacement from the square planar surrounding with a non-$d^0$-configuration leads to the breaking of an inversion center similar to ref. \cite{li2018new}. Further, Fe-spins interact antiferromagnetically with both Mn-, and Ni-spins. Consequently, ferromagnetic (FM) interactions are found between Mn and Ni spins. These FM interactions govern a net 5.00 $\mu_B$/f.u. magnetization into the system (TABLE \ref{tab:2}). Since the nearest neighbor distances reduce, we can expect even stronger magnetic exchange interactions as compared to CaMnTi$_2$O$_6$. The magnitudes of Fe-moments (4.21 $\mu_B$/Fe) of SP and TH FeO$_4$ are found to be almost similar.
\par
The lowest magnetic configuration of La$^{3+}$Ti$^{4+}$Mn$^{3+}$Ni$^{2+}$O$_6$ is found to be A-type antiferromagnetic (AFM) ordering as described in Figure \ref{DOS}(c), followed by ferromagnetic (FM) configuration. The corresponding electronic structure is shown in Figure \ref{DOS}(d). The AFM ordering with the absence of an inversion center indicates that this compound is important for multifunctional properties. Other collinear spin configurations (Figure \ref{TiMagnetism}) are found to be stable within an enhanced energy window ($\sim$ 40 meV) as compared to ref. \cite{gou2017site} with reference to A-type AFM-ordering. This indicates that we may achieve strong magnetic exchange interactions for LaTiMnNiO$_6$ as compared to CaMnTi$_2$O$_6$.
\par
\begin{figure}
\centering

\includegraphics[width=\linewidth]{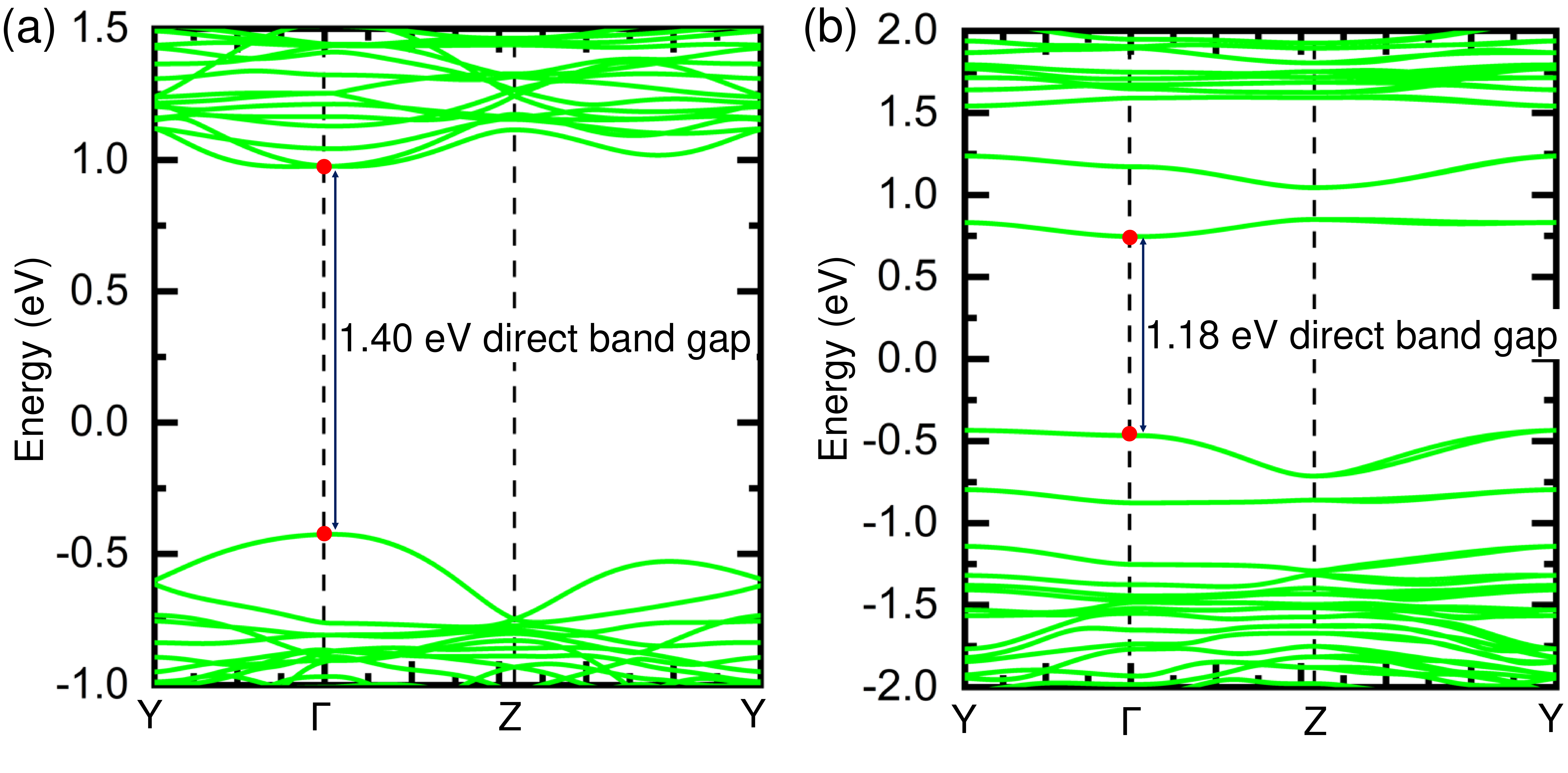}\vspace{-0pt}
\caption {(Color online) Calculated electronic band structure from GGA+U calculation for (a) LaFeMnNiO$_6$ and (b) LaTiMnNiO$_6$ DDPOs.} 
\label{BandStructures}
\end{figure}
\begin{figure}
\centering
\includegraphics[width=\linewidth]{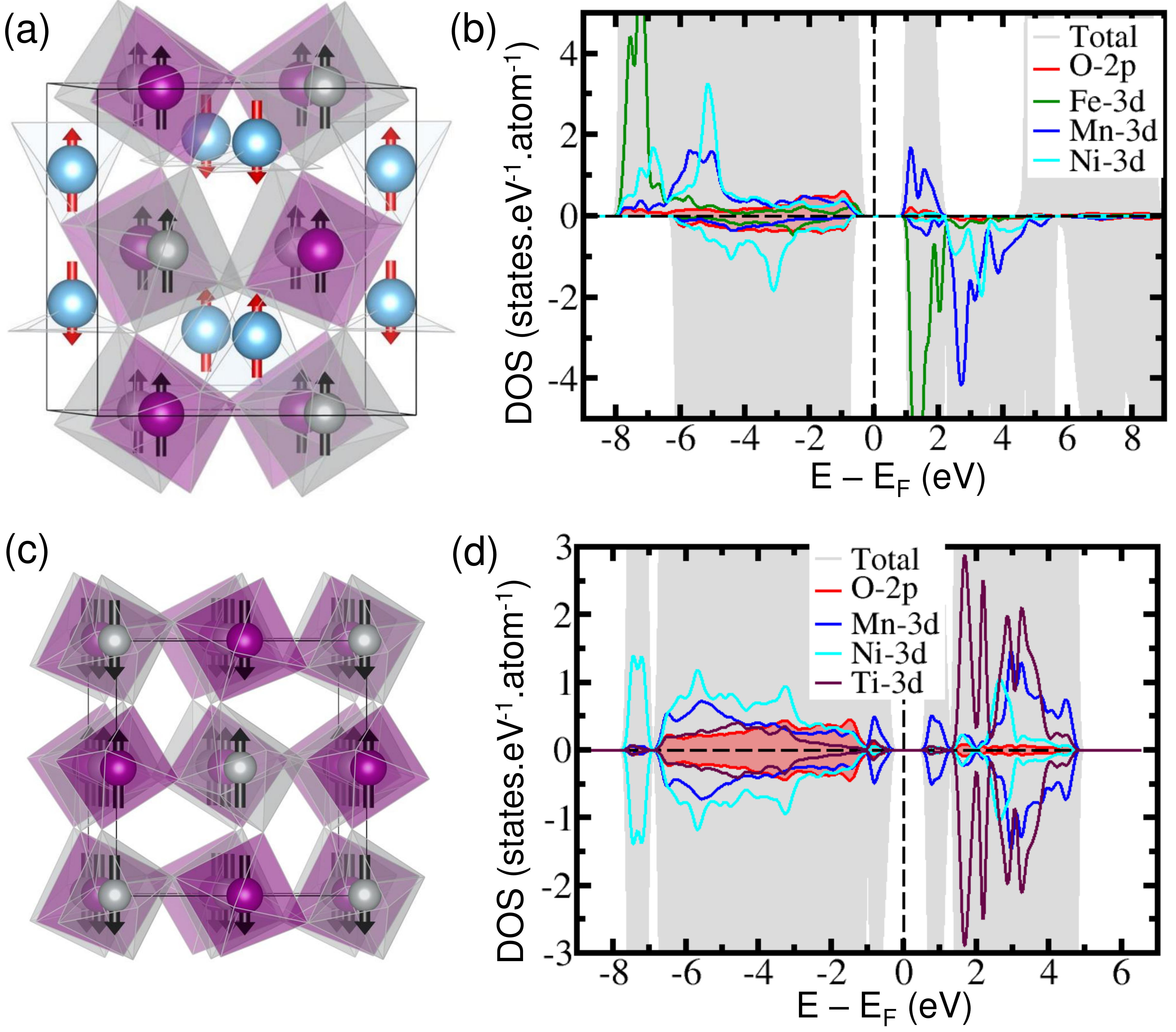}\vspace{-0pt}
\caption {(Color online) Ground state spin configuration and corresponding electronic structure of LaFeMnNiO$_6$ (top panel) and LaTiMnNiO$_6$ (bottom panel). The non-magnetic atoms are not shown for clarity. The Fe-, Mn-, and Ni-atoms are described by blue, magenta, and silver balls respectively.} 
\label{DOS}
\end{figure}

The magnetic polar behavior of LFMNO and LTMNO makes them superior compounds in the family. Furthermore, we identify LFMNO, and LTMNO as direct bandgap semiconductors with energy gaps of 1.40, and 1.18 eV respectively and presented in Figure \ref{BandStructures}. This may be facilited them for visible light absorption. Density of states (DOS) analysis of LFMNO reveals that a local moment of 4.21 $\mu_B$/Fe with filled $d$-orbitals in the up spin channel (USC) suggests a nominal charge state of Fe$^{3+}$ ($t_{2g}^3e_g^2$) is in a high spin state and is shown in Figure \ref{DOS} (b). The octahedral environment of Mn-, and Ni-atom show a $t_{2g}$, and $e_{g}$ crystal field splitting. Ni-$t_{2g}$, and Ni-$e_{g}$ bands in the USC lie between -8 eV energy and the Fermi energy (E$_F$) and show a strong hybridization with Mn-$d$ and O-$p$. While Ni-$t_{2g}$ bands in the down spin channel (DSC) are localized between O-$p$ and the Fermi level. The $e_{g}$ levels in the DSC are located above E$_F$. This DOS along a local moment of 1.71 $\mu_B$/Ni indicates a nominal charge state of Ni$^{2+}$ ($t_{2g}^6e_g^2$). The filled Mn-$t_{2g}$ bands are located between Ni-$t_{2g}$ and Ni-$e_{g}$ levels in the USC. Mn-$t_{2g}$ in the DSC is totally empty. While Mn-$e_{g}$ in both the spin channels are found above the Fermi level and an insulting phase is obtained analogous to ref. \cite{PRL2008}. This suggests a nominal charge state of Mn$^{4+}$ ($t_{2g}^3e_g^0$) with local moment of 3.16 $\mu_B$/Mn.    
\begin{figure}
\centering

\includegraphics[height= 5.6 cm]{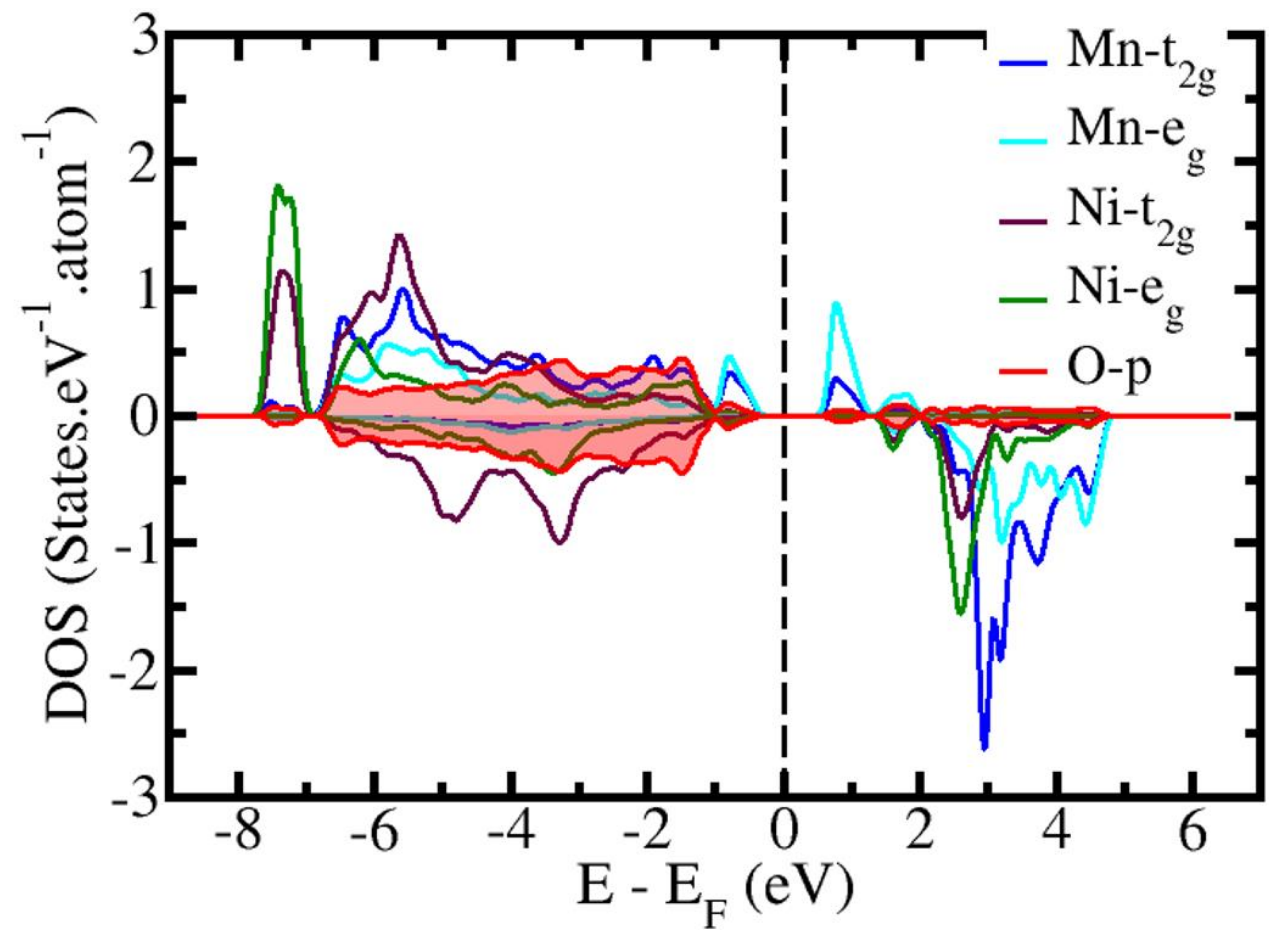}\vspace{-0pt}
\caption {(Color online) Orbital resolved electronic structure from GGA+U calculation for LaTiMnNiO$_6$ DDPO.} 
\label{TiDOS}
\end{figure}
\par
Investigation on the partial DOS of LTMNO, provides us with Ti-$d$ states to be mostly empty, suggesting its $d^0$-configuration (Ti$^{4+}$) and is shown in Figure \ref{DOS} (d). Due to octahedral surrounding in MnO$_6$, and NiO$_6$ the $d$-orbitals split up into $t_{2g}$, and $e_{g}$ levels similar to LFMNO. As discussed earlier the Mn-Ni interaction in LTMNO between layers (A-type AFM ordering) is antiferromagnetic and hence the DOS is identical in both the spin-channels. However, if we consider any FM pair of Mn- and Ni-atoms, we find a similar electronic structure for the Ni$^{2+}$ ($t_{2g}^6e_g^2$) and is shown in Figure \ref{TiDOS}. While the filled Mn-$t_{2g}$ bands are found between Ni-$t_{2g}$ and Ni-$e_{g}$ levels in the USC. But, Mn-$e_{g}$ lies below and above E$_F$ in the same USC. In the DSC both Mn-$t_{2g}$ and Mn-$e_{g}$ are located above Fermi energy leading to an insulating solution. This suggests a nominal charge state of Mn$^{3+}$ ($t_{2g}^3e_g^1$) with local moment of 3.82 $\mu_B$/Mn.  
\begin{figure*}
\centering
\includegraphics[height= 5.5 cm]{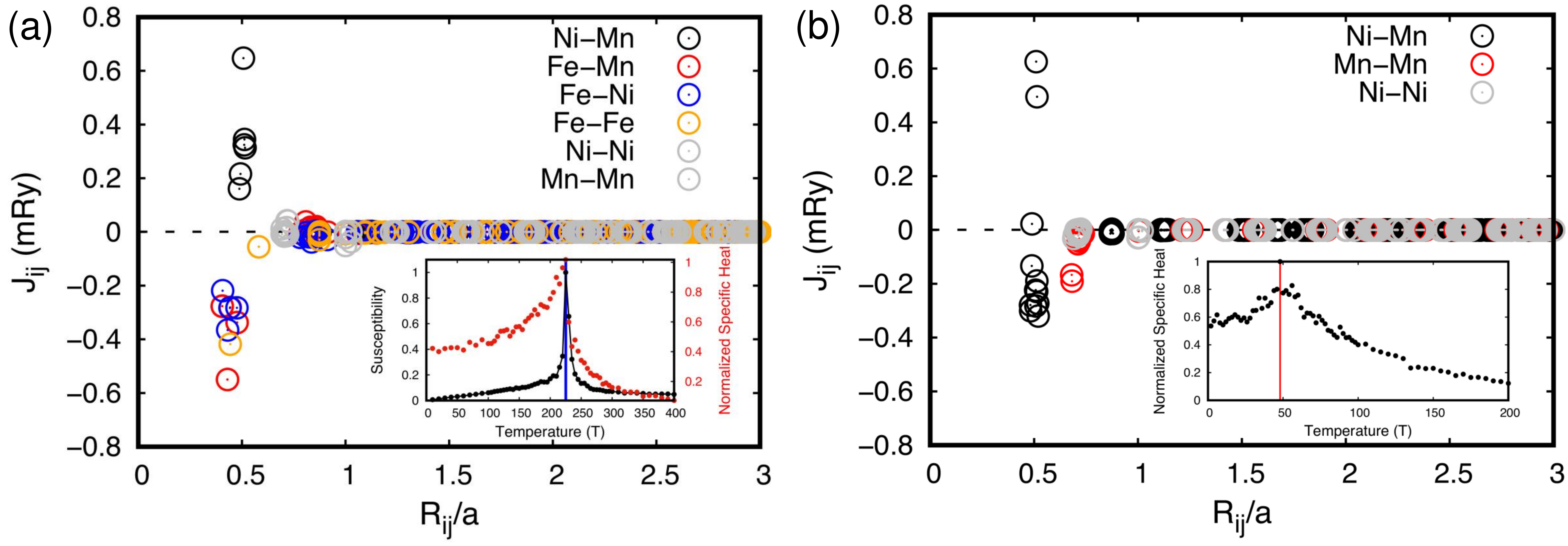}\vspace{-0pt}
\caption {(Color online) (a) Calculated inter-site exchange parameters for each $TM$-$TM$ pair as a function distance for LFMNO. The positive and negative values represent ferromagnetic and antiferromagnetic couplings, respectively. The inset shows the magnetization (black circles) and the normalized specific heat (red circles) as a function of temperature, calculated from classical Monte Carlo simulations. A blue vertical line indicates the magnetization transition temperature. (b) Calculated inter-site exchange parameters for each $TM$-$TM$ pair as a function distance for LTMNO. The inset shows the normalized specific heat as a function of temperature, calculated from classical Monte Carlo simulations. A red vertical line indicates the magnetization transition temperature.}
\label{MonteCarlo}
\end{figure*}
\subsection{Monte Carlo Simulations}
The calculated inter-site exchange parameters are shown in Figure \ref{MonteCarlo} (a). First, the magnetic couplings decrease rapidly as the TM-TM distance increases; only the contributions from the first few nearest neighbours play an important role in determining the magnetic ground state. Second, among the couplings between the three transition metal elements, the one between Ni and Mn is ferromagnetic, while those for the other two pairs contribute to antiferromagnetic interaction. In addition, different from the nearest Ni-Ni and Mn-Mn pairs, which are separated by nearly a unit cell's distance, the two Fe-Fe pair distances are relatively small and, therefore, exhibit noticeable antiferromagnetic coupling. This is perfectly consistent with the ferrimagnetic ground state configuration obtained from our total energy calculations. 
\par
With an average Ni-Mn ferromagnetic coupling of 0.33 mRy, and the Fe-Mn and Fe-Ni antiferromagnetic couplings of -0.35 mRy and -0.28 mRy, respectively, in addition to one strong and one weak antiferromagnetic Fe-Fe coupling of -0.42 mRy and -0.06 mRy, the system exhibits a remarkable magnetic transition temperature of 225 K, as shown in the inset of FIG. 4. 
In the case of LTMNO, normalized specific heat as a function of temperature exhibits a magnetic transition temperature of 48 K calculated from classical Monte Carlo simulations and is shown in Figure \ref{MonteCarlo} (b). 

\subsection{Computation of Formation Energies}
To identify the correct reaction sequence, we follow a theoretical framework suggested by Akbarzadeh \textit{et al.} ~\cite{Akbarzadeh2007} and made appropriate modifications to suit our problems~\cite{CmemMat2021, DuoPRM}.
We consider all possible stable oxides from the Materials Project ~\cite{jain2013commentary} that fall on the energy hull and are listed in TABLE \ref{tab:1}. The reaction energy is calculated by the following expression:
 \begin{center} \label{eq:1}
\begin{equation}
   G = \sum_{i} x_{i} F_{i}
\end{equation}
\end{center}
Where $G$ is the total reaction energy of the reaction sequences, $i$ includes a set of all possible stable oxides, $F$ is the free energy (at T = 0 K) of the $i$th compound, and $x_i$ (unknown) is the variable molar fraction of the $i$th compound at a given composition. To get the minimum energy reaction, we minimized equation 1 with respect to the molar fraction $x_{i}$, with mass conservation constraints such as
 \begin{center} \label{eq:2}
\begin{equation}
   f_{s} = \sum_{i} x_{i} n^s_{i} = Constant 
\end{equation}
\end{center}
Where $n^s_{i}$ is the number of ions of species $s$ in the $i$th compound per formula unit and $f_s$ is the molar fraction of the individual species $s$ (s = La, Ti, Fe, Mn, Ni, and O). To apply the above formalism, we considered a set of all possible stable compounds from the Materials Project database ~\cite{jain2013commentary} that contains La, Ti, Fe, Mn, Ni, and O as the constituents of the DDPOs. We calculate the free energy of these compounds using the same parameters that we used to calculate the energies of DDPOs. Using these calculated free energies, we minimized the linear equation 1 using a linear programming solver and calculated the formation energies with respect to these minima
energy reactions.
\begin{table}[b]
\begin{center}
\caption{\label{tab:1} List of stable oxides considered for computing the formation energies of the double-double
perovskites.}
\begin{ruledtabular}
\begin{tabular}{cc|cc}
Stable Oxides & Space group & Stable Oxides & Space group\\ 
\hline
La$_2$O$_3$ & $Ia-3$ & TiNiO$_3$ & $R-3$ \\
La$_2$TiO$_5$ & $Pnma$ & TiMnO$_3$ & $R-3$ \\
La$_2$Ti$_2$O$_5$ & $P2_1$ & TiMn$_2$O$_4$ & $P4_32_2$\\
Ti$_2$O & $P-3m_1$ & Mn$_2$O$_3$ & $Pbca$\\
Ti$_2$O$_3$& $R-3c$ & MnO$_2$& $I4/m$ \\
Ti$_3$O$_5$& $C2/m$ & MnO & $Fm-3m$\\
Ti$_6$O & $P-31c$ & LaNiO$_3$ & $R-3c$\\
TiO & $P-62m$ & MnNiO$_3$ & $R-3$ \\
NiO & $Fm-3m$ & Fe$_2$NiO$_4$ & $Imma$ \\
Ni$_3$O$_4$ & $Cmmm$ & Mn$_3$O$_4$ & $I41/amd$\\
LaFeO$_3$ & $R-3c$ &  FeO & $C2/m$\\
FeO & $C2/m$ & Fe$_2$O$_3$ & $R-3c$\\
Ti$_3$O & $P-31c$ & TiO$_2$ & $C2/m$\\

\end{tabular}
\end{ruledtabular}
\end{center}
\end{table}
The thermodynamic stability is one of the basic requirements to synthesize a compound for magnetization into the systempractical applications. We, therefore, examine the formation energies of LFMNO and LTMNO DDPOs. We calculate the formation energies for these systems by considering the decomposition reaction of DDPOs via the most probable reaction pathways by utilizing a linear programming problem clubbed with the grand canonical method~\cite{CmemMat2021}. The detailed methodology is provided in the method section. We consider the stable oxides that were reported in the inorganic crystal structure database (ICSD) ~\cite{ruhl2019inorganic}, and in the materials project ~\cite{jain2013commentary} in our study. The minimum energy decomposition paths for these compounds are:  
\begin{equation}
\begin{split}
\Delta_fE_5[LaFeMnNiO_{6}] = E[LaFeMnNiO_{6}]\\-(\frac{3}{6}E[La_{2}O_{3}]+ \frac{3}{6}E[Fe_{2}O_{3}]+\frac{1}{6}E[Mn_{3}O_{4}]\\+\frac{3}{6}E[MnO_{2}]+\frac{2}{6}E[Ni_{3}O_{4}])
\end{split}
\end{equation}
\begin{equation}
\begin{split}
\Delta_fE_1[LaTiMnNiO_{6}] = E[LaTiMnNiO_{6}]\\-(\frac{3}{6}E[La_{2}Ti_{2}O_{7}]+ \frac{2}{6}E[Mn_{3}O_{4}]+\frac{3}{6}E[NiO]\\+\frac{1}{6}E[Ni_{3}O_{4}])
\end{split}
\end{equation}
Implementing these equations, the formation energies of LFMNO and LTMNO are found to be 1.25 eV and 0.89 eV respectively. This is in agreement with the experimental observations as the DDPOs are synthesized under high pressure ($\sim$ 10-15 GPa) and high-temperature ($\sim$ 1200-1700 $^o$C)\cite{mcnally2017complex, ji2023cafefenbo}.
\section{CONCLUSION}
In summary, our investigation based on DFT calculations and symmetry analysis reveals the presence of direct gap semiconductors, with GGA+U forbidden energy values of 1.40 eV and 1.18 eV for LFMNO and LTMNO, respectively, suggesting potential suitability for visible light absorption. We elucidate the origins of these magnetic polar semiconductors. The observed long-range ferri-magnetic ordering in LFMNO is attributed to superexchange interactions. Through Monte Carlo simulations, we determine magnetic transition temperatures of 225K and 48K for LFMNO and LTMNO, respectively, significantly higher than the approximate 10 K observed in CaMnTi$_2$O$_6$. The calculated spontaneous polarization we report as
20.0 and 21.8 $\mu C$/cm$^2$ for LFMNO and LTMNO, respectively. In conclusion, both LFMNO and LTMNO emerge as promising multiferroics, featuring magnetic transition metals at B- and B$^\prime$-sublattices, in addition to TM at the A$^\prime$-site.
\section{ACKNOWLEDGMENTS}
M.S. acknowledges INSPIRE division, Department of Science and Technology, New Delhi-110 016, Government of India, for his fellowship [IF170335].
S.G.  acknowledges the DST-SERB Core Research Grant (File No. CRG/2018/001728) for funding. D.W. acknowledges financial support from the Science and Technology Development Fund from Macau SAR (0062/2023/ITP2) and the Macao Polytechnic University (RP/FCA-03/2023).
\bibliographystyle{apsrev4-2}
%

\end{document}